\documentclass[twocolumn,prd,amsmath,amsfonts,nofootinbib]{revtex4}
\newcommand{\half}{{\tfrac{1}{2}}}
\newcommand{\gamsem}{\breve{{;}}}
\newcommand{\gamgra}{\breve{\nabla}}
\newcommand{\ud}{\mathrm{d}}
\begin{document}
\title{The physical significance of the Babak-Grishchuk \\ gravitational energy-momentum tensor}
\author{Luke M. Butcher}
\email{l.butcher@mrao.cam.ac.uk}
\author{Anthony Lasenby}
\author{Michael Hobson}
\affiliation{Astrophysics Group, Cavendish Laboratory, Madingley Road, Cambridge CB3 0HE, UK.}
\date{\today}

\begin{abstract}
We examine the claim of Babak and Grishchuk \cite{BG} to have solved the problem of localising the energy and momentum of the gravitational field. After summarising Grishchuk's flat-space formulation of gravity, we demonstrate its equivalence to General Relativity at the level of the action. Two important transformations are described (diffeomorphisms applied to all fields, and diffeomorphisms applied to the flat-space metric alone) and we argue that both should be considered gauge transformations: they alter the mathematical representation of a physical system, but not the system itself. By examining the transformation properties of the Babak-Grishchuk gravitational energy-momentum tensor under these gauge transformations (infinitesimal and finite) we conclude that this object has no physical significance.
\end{abstract}
\maketitle
\section{Introduction}
Despite the central role played by the energy-momentum tensor of matter in General Relativity, there is no widely accepted way to localise the energy and momentum of the gravitational field itself. In the place of a genuine solution to this problem, we are forced to make do with an over-abundance of energy-momentum pseudotensors, objects designed to display some or other property befitting a measure of gravitational energy-momentum, but whose coordinate dependence renders them of little physical significance beyond giving the correct integrals at infinity in asymptotically flat spacetimes. Even for weak gravitational waves, the best measures at our disposal only become meaningful once we have averaged over many wavelengths.

The canonical response to the gravitational energy-momentum problem is to dismiss it as ``looking for the right answer to the wrong question''\cite{MTW}; but while the well-known argument presented by Misner, Thorne and Wheeler is certainly compelling, it is far from watertight. They remind us that the equivalence principle ensures that all ``gravitational fields'' $\Gamma^{\alpha}_{\phantom{\alpha}\beta \gamma}$ can be made to vanish at a point by a suitable choice of coordinates, and conclude that because gravity is locally zero, there can be no energy density associated with it. However, this argument fails to consider tensors containing second derivatives of the metric, which unlike $\Gamma^{\alpha}_{\phantom{\alpha}\beta \gamma}$ cannot be made to vanish by choice of coordinates, and really do reflect the local curvature of spacetime: for example, the Riemann tensor can be used to construct objects such as the Bel-Robinson tensor \cite{Bel}. Misner, Thorne and Wheeler also point out that, while the matter energy-momentum tensor derives its physical significance by curving space, a similar tensor for gravity would not be a source term for the field equations. However, this stance is based around a prejudice for writing the Einstein field equations as $G^{ab} = \kappa T^{ab}$ with gravity on the left and matter on the right; there is nothing to stop us splitting up $G^{ab}$ in a covariant fashion, grouping one part with $T^{ab}$, and interpreting \emph{this} as the total energy-momentum source, taking the remainder of $G^{ab}$ to be the gravitational `response'. Despite these reservations, the argument in \cite{MTW} remains vindicated as yet by the failure of these escape-routes to yield anything which can be physically interpreted as an energy-momentum tensor.

It might appear that the only straightforward solution to the problem is to extend the definition of the matter energy-momentum tensor $T^{ab}$ (a functional derivative of the matter Lagrangian with respect to the metric) to the gravitational field, and conclude that the gravitational energy-momentum tensor is $-G^{ab}/\kappa$, where $\kappa= 8\pi G/c^4 $. The Einstein field equations could then be interpreted as a constraint that everywhere sets to zero the sum of gravitational and matter energy-momentum. While one might claim this simple idea conveys some important physical insight, it suffers from numerous problems. Firstly, $-G^{ab}/\kappa$ lacks the analytical power one expects from an energy-momentum tensor: the ability to split the set of all physical systems at a particular time into classes of \emph{different} total energy and momenta, so that conservation laws alone can reveal that two particular systems could never be part of the same spacetime. Secondly, it leads us to conclude that the gravitational field only has energy where matter is also present, precluding the use of this prescription to describe the energetics of gravitational waves, or define a gravitational tension in the vacuum between massive bodies. Thirdly, the energy-momentum tensors for gravity and matter are conserved \emph{separately} ($\nabla_a G^{ab}=0$ and $\nabla_a T^{ab}=0$) so that although there is a delicate balance that keeps their sum zero, it is not the case that energy or momentum simply `flows' between gravity and matter, as $\nabla_a (T^{ab} - G^{ab}/\kappa)=0$ \emph{alone} would imply. Lastly, we note that the conservation law $\nabla_aG^{ab}=0$ actually tells us nothing at all about the gravitational field; it is satisfied identically, without any need for the equations of motion to hold. Because of these drawbacks, if we are to regard $-G^{ab}/\kappa$ as a solution to the gravitational energy-momentum problem, we consider it rather a trivial one. Clearly, the reason for this triviality is that we have over-worked the metric: we cannot use the functional derivative with respect to a dynamical field as a way of defining the energy-momentum tensor for that same field, as we will only end up writing down the equations of motion twice. This line of reasoning leads us to consider that one method of attack for this problem may be to separate the two roles played by $g^{ab}$ in General Relativity, that of dynamic field and spacetime metric. 

In \cite{G}, Grishchuk develops a ``field-theoretical'' approach to gravitation, which expresses the physical content of General Relativity (GR) in terms of a dynamical symmetric tensor field in flat Minkowski spacetime. Although this formulation has been carefully designed to agree with the empirical predictions of GR, in \cite{BG} Babak and Grishchuk claim that the flat-space approach allows them to define a unique, symmetric, and non-trivial energy-momentum tensor for the gravitational field. The major purpose of this paper is to examine the extent to which this tensor is physically meaningful.

\section{Flat-Space Gravitation}
Babak and Grishchuk represent gravitation as the theory of a dynamical symmetric tensor field $h^{ab}$ defined over a four-dimensional manifold $\mathcal{M}$ with a (non-dynamical) flat Lorentzian metric $\gamma^{ab}$. Translation between this picture and the dynamical metric $g^{ab}$ of GR can be achieved using the following relation:
\begin{eqnarray} \label{split}
\sqrt{-g}g^{ab}=\sqrt{-\gamma}(\gamma^{ab} + h^{ab}),
\end{eqnarray}
where $g=1/\textrm{det}(g^{\alpha\beta})$ and $\gamma=1/\textrm{det}(\gamma^{\alpha\beta})$. It should be emphasised that Babak and Grishchuk consider this relation to be the \emph{definition} of $g^{ab}$, a tensor to which they assign no particular fundamental or geometric significance.\footnote{Of course, because Babak and Grishchuk insist that this viewpoint does not contradict the predictions of General Relativity, effects that are traditionally deemed the result of spacetime geometry (proper lengths of coordinate displacements, rates of clocks, geodesic deviation, etc.) will be viewed as arising from a 4-force that matter feels in response to the presence of $h^{ab}$; see \cite{G} for details. The correspondence with GR inevitably means that predictions of this nature can always be expressed in terms of $g^{ab}$ alone.} Accordingly, they use $\gamma^{ab}$, rather than $g^{ab}$, to raise and lower tensor indices\footnote{A singular exception is made for $g^{ab}$: it is assigned the `lowered' form $g_{ab}=(g^{ab})^{-1}$ to coincide with the GR definition.}, and define a (torsion-free) covariant derivative $\gamgra_a$ (denoted by indices following ``$ \ {\gamsem} \ $''\footnote{This notation differs from \cite{BG}: where they write $\nabla$ and ``$ \ ; \ $'', we write $\gamgra$ and ``$ \ \gamsem \ $''.} and with Christoffel symbols $C^a_{\phantom{a}bc}$) by $\gamgra_c\gamma^{ab}=0$. 
As $\gamma^{ab}$ is flat,
\begin{eqnarray}\label{flat}
\breve{R}^a_{\phantom{a}bcd} \equiv C^a_{\phantom{a}bd{,}c} - C^a_{\phantom{a}bc{,}d} + C^e_{\phantom{e}bd}C^a_{\phantom{a}ec} - C^e_{\phantom{e}bc}C^a_{\phantom{a}ed} = 0,
\end{eqnarray}
and $\gamgra_a$ derivatives commute. This contrasts to the usual (GR) covariant derivative $\nabla_a$, denoted by indices following ``$ \ ; \ $'', defined by $\nabla_cg^{ab}=0$, and with curvature tensor $R^a_{\phantom{a}bcd} \ne 0$ in general.\footnote{There is no contradiction in being able to define two different covariant derivatives on a manifold. Because both have been defined by a tensor equation (without any reference to coordinate systems) they must both produce genuine (abstract) tensor indices  ${{;}a}$  and  ${{\gamsem}a}$. The significance of the standard covariant derivative (in GR) is not just that it is covariant, but that it expresses the Equivalence Principle: in a system of local inertial coordinates $\{x^\alpha\}$ such that $g^{\alpha\beta} = \eta^{\alpha\beta} + O(x^2)$ near some point $p$, the Christoffel symbols for the $\nabla_a$ derivative vanish and we find that (at $p$) $\nabla_a=\partial_a$, the ordinary derivative of these coordinates. Thus $\nabla_c g^{ab}=0$ picks out the coordinate independent derivative operator which coincides with local inertial coordinate derivatives. In contrast, a coordinate system $\{y^\alpha\}$ for which $\gamgra_a =\partial_a$ at $p$ will not necessarily have $g^{\alpha\beta} = \eta^{\alpha\beta} + O(y^2)$ there; however, as the flat-space picture eschews the geometric interpretation of $g^{ab}$, we can avoid assigning much significance to this point.}

To ensure that $h^{ab}$ obeys an equation of motion consistent with Einstein's field equations, its dynamics are determined by an action $S$ that is equivalent to the Einstein-Hilbert action. Specifically, in \cite{BG} Babak and Grishchuk use the action
\begin{eqnarray}
S &=& \tfrac{-1}{2\kappa}\int\sqrt{-\gamma} \Big[h^{ab}_{\phantom{ab}\gamsem c}P^c_{\phantom{c}ab} \nonumber \\
&& \qquad {}- (\gamma^{ab} + h^{ab})(P^c_{\phantom{c}ad}P^d_{\phantom{d}bc} - \tfrac{1}{3}P_a P_b)\Big] \ud^4\! x,
\end{eqnarray}
where $P^a_{\phantom{a}bc}$ and $P_a$ are functions of $h^{ab}$, $\gamma^{ab}$ and $h^{ab}_{\phantom{ab}\gamsem c}$ given in their paper. If we add to the Lagrangian the following surface term:
\begin{eqnarray}
\mathcal{L}_\textrm{surface} &=&\tfrac{1}{2\kappa}\left[\sqrt{-\gamma}(\gamma^{ab} +h^{ab})P^c_{\phantom{c}ab}\right]_{{,}c} \nonumber\\
&=& \tfrac{1}{2\kappa}\left[\sqrt{-\gamma}(\gamma^{ab} +h^{ab})P^c_{\phantom{c}ab}\right]_{\gamsem c},\nonumber
\end{eqnarray}
then, applying the flatness condition (\ref{flat}) to equation (53) of \cite{BG}, we see that
\begin{eqnarray}
S + S_{\textrm{surface}} & = & \tfrac{1}{2\kappa}\int\sqrt{-\gamma} (\gamma^{ab} + h^{ab})  \nonumber \\
&& \qquad \times\left(P^c_{\phantom{c}ab\gamsem c} + P^c_{\phantom{c}ad}P^d_{\phantom{d}bc} - \tfrac{1}{3}P_a P_b\right) \ud^4\! x \nonumber \\
& = & \tfrac{-1}{2\kappa}\int\sqrt{-g} g^{ab} R_{ab} \ud^4\! x \nonumber \\
& = & S_{\textrm{EH}},
\end{eqnarray}
the Einstein-Hilbert action. Minimising $S$ with respect to a variation in $h^{ab}$, we have
\begin{eqnarray}\label{eom}
\frac{\delta S}{\delta h^{ab}} = 0 \quad \Rightarrow \quad \frac{\delta S_{\textrm{EH}}}{\delta g^{cd}}\left(\frac{\partial g^{cd}}{\partial h^{ab}}\right)_{\!\gamma} = 0,
\end{eqnarray}
where the subscript $\gamma$ indicates that $\gamma^{ab}$ has been held constant. As an inverse of $\left(\frac{\partial g^{cd}}{\partial h^{ab}}\right)_{\!\gamma}$ exists, namely
\begin{eqnarray}
\left(\frac{\partial h^{ab}}{\partial g^{cd}}\right)_{\!\gamma} & = & \frac{1}{\sqrt{-\gamma}}\frac{\partial\sqrt{-g}g^{ab}}{\partial g^{cd}} \nonumber \\ 
& = & \frac{\sqrt{-g}}{\sqrt{-\gamma}}\left( 2\delta^{(a}_{\phantom{(a}c} \delta^{b)}_{\phantom{b)}d} - \half g^{ab}g_{cd} \right),\nonumber
\end{eqnarray}
the equations of motion (\ref{eom}) are equivalent to the Einstein Field Equations:
\begin{eqnarray}\label{einst}
\frac{\delta S_{\textrm{EH}}}{\delta g^{ab}} = 0.
\end{eqnarray}

As presented in \cite{G}, the original motivation for this flat-space picture is that it allows physicists to study and predict gravitational phenomena in a framework that is free of the conceptual baggage of differential geometry, and has more in common with the language of particle physics and classical electrodynamics. However, the work presented in \cite{BG} elevates this framework beyond the status of a `linguistic trick', as the metric $\gamma^{ab}$ allows one to define the ``metrical energy-momentum tensor'' according to
\begin{eqnarray}\label{tmet}
\overset{\ m}{t}\!{}^{ab} &\equiv& \frac{-2}{\sqrt{-\gamma}}\frac{\delta\mathcal{L}}{\delta\gamma_{ab}} \nonumber \\ 
&\equiv& \frac{-2}{\sqrt{-\gamma}}\left(\frac{\partial\mathcal{L}}{\partial\gamma_{ab}} - \partial_c\left(\frac{\partial\mathcal{L}}{\partial\gamma_{ab,c}}\right)\right).
\end{eqnarray}
From this, a unique gravitational energy-momentum tensor $t^{ab}$ can be constructed that is symmetric, free of second derivatives, and conserved by the equations of motion: see equation (65) of \cite{BG}. Having made this identification, the field equations (\ref{eom}) take on the simple form\footnote{This result is a corrected version of equation (78) of \cite{BG}; it is easy to see that the original equation lacks a factor of $-1/\gamma$ by comparing it to the preceding equation in that paper.} 
\begin{eqnarray}\label{tab}
\kappa t^{ab}= \left[\frac{g}{2\gamma}(g^{ab}g^{cd} - g^{ac}g^{bd})\right]_{\gamsem c \gamsem d}.
\end{eqnarray}
Although this equation does not \emph{define} the energy-momentum tensor, it provides us with a simple method for calculating $t^{ab}$, given the gravitational field.

\section{The Physical Content of $t^{ab}$}
We cannot fault Grishchuk's formulation of gravitational dynamics within the realm of General Relativity, as agreement over predictions of `geometrical phenomena' (as they would be interpreted in GR) has been achieved by design.\footnote{Of course, one may still wish to attack the \emph{aesthetics} of a framework which, from the GR viewpoint, appears to obscure the geometric nature of gravity, and replaces the Equivalence Principle with a seemingly arbitrary coupling between $h^{ab}$, $\gamma^{ab}$ and matter. However, the potential for a greater understanding of the local energy-momentum content of the gravitational field should be enough to temporarily assuage these objections.} However, in comparison with General Relativity, the flat-space theory possesses additional mathematical structure: two tensors $h^{ab}$ and $\gamma^{ab}$ fulfil the role played by $g^{ab}$ alone. This extra structure endows the flat-space theory with an increased range of expression, making possible the definition of tensors that cannot be constructed within the framework of GR. As we shall show, the gravitational energy-momentum tensor is one of these `non-GR' quantities.\footnote{This statement might appear obvious due to the use of $\gamma^{ab}$ in (\ref{tmet}), or the presence $h^{ab}_{\phantom{ab}\gamsem c}$ in the definition of $t^{ab}$ (equation (65) of \cite{BG}). However, a tensor defined in terms of $\gamma^{ab}$, $h^{ab}$, and $\gamgra_a$ may also be expressible in GR, e.g.~$P^c_{\phantom{c}ab\gamsem c} + P^c_{\phantom{c}ad}P^d_{\phantom{d}bc} - \tfrac{1}{3}P_a P_b = - R_{ab}[g]$.} We investigate here whether $t^{ab}$ (or any non-GR quantity) can be physically significant, or whether it can only ever be interpreted as an artefact of the mathematics.   

\subsection{Gauge transformations}
Besides allowing us to interpret gravity as a force-field on flat space, the presence of $\gamma^{ab}$ has had the important side-effect of increasing the space of gauge transformations of the theory. The core reason for this is that the flatness constraint (\ref{flat}) is not enough to define a unique $\gamma^{ab}$ for a given $g^{ab}$, a tensor which, through the correspondence with GR, can be used \emph{alone} to construct the observable predictions of the theory. In this section we examine two transformations and justify their status as gauge transformations, i.e.~that they alter the mathematical representation of a physical system, but not the system itself.

\subsubsection{Diffeomorphism gauge transformations}
Given a diffeomorphism $\phi{:} \ \mathcal{M} \to \mathcal{M}$, we transform all tensor fields $X^{a\dots}_{\phantom{a\dots}b\dots}$ according to
\begin{eqnarray}\label{dif}
X^{a\dots}_{\phantom{a\dots}b\dots} \to (\phi^*X)^{a\dots}_{\phantom{a\dots}b\dots},
\end{eqnarray}
where the action of $\phi^*$ on $X$ is defined in the standard way by the action of the pullback of $\phi$ (and the pushforward of $\phi^{-1}$) on the dual-vector (and vector) arguments of $X$; see \cite{wald} for details. Although, as written, this transformation cannot be the result of a change of coordinate system\footnote{In this paper we use the \emph{abstract index notation} developed by Penrose and Rindler \cite{pen}, so that the Roman indices of $\gamma^{ab}$ indicate the tensor `slots' of the metric, and do not refer to components of the tensor in any coordinate system. Thus the `effect' of a coordinate transformations is completely invisible to a tensor equation notated with abstract indices. To notate the matrix of components of a tensor such as $\gamma^{ab}$ in coordinates $\{x^\alpha\}$ we use Greek indices: $\gamma^{\alpha\beta}\equiv \gamma^{ab}(\ud x^\alpha)_a(\ud x^\beta)_b$.}, it transforms the \emph{components} $X^{\alpha\dots}_{\phantom{\alpha\dots}\beta\dots}$ in a typographically identical manner to that of a coordinate change. More precisely, the components of $\phi^*X$ at $\phi(p)$ in a coordinate system $\{x^\alpha \}$ will be equal to the components of $X$ at $p$ in coordinates $\{y^\alpha\}$ where $y^\alpha(q) = x^\alpha(\phi^{-1}(q))$. As such, if we had chosen to represent all our tensor equations in terms of components in some coordinate system, it would be impossible to tell (from the transformation law alone) whether we had performed the diffeomorphism (\ref{dif}) or simply changed coordinates. Therefore, because the physical content of a tensor field's \emph{components} cannot depend on which coordinate system it is expressed in, so the physical content of \emph{tensor fields} cannot depend on the action of (\ref{dif}). Thus, just as in General Relativity, we find that Grishchuk's formulation contains the group of diffeomorphisms $\phi{:} \ \mathcal{M} \to \mathcal{M}$  as a gauge freedom.

\subsubsection{The $\gamma$-transformation} 
Besides the diffeomorphism gauge transformation (DGT), it is also possible to use a diffeomorphism $\phi{:} \ \mathcal{M} \to \mathcal{M}$ to define a transformation that reflects the range of flat-metrics $\gamma^{ab}$, and gravitational fields $h^{ab}$, consistent with a particular $g^{ab}$; we apply the diffeomorphism to $\gamma^{ab}$ alone, and demand that $h^{ab}$ compensate in such a way that $g^{ab}$ remains unchanged:
\begin{eqnarray}\label{gen}
\gamma^{ab}&\to& (\phi^*\gamma)^{ab}, \nonumber \\
h^{ab} &\to& {h'}^{ab} = \tfrac{\sqrt{-\gamma}}{\sqrt{-\phi^*\gamma}}\left( \gamma^{ab} + h^{ab} \right) - (\phi^*\gamma)^{ab}, \nonumber\\
\Rightarrow g^{ab} &\to& g^{ab}.
\end{eqnarray}
To be consistent with the field equations, if we are to include matter fields $M^{a\dots}_{\phantom{a\dots}b\dots}$ in the theory, we must make them similarly invariant:
\begin{eqnarray}
\quad M^{a\dots}_{\phantom{a\dots}b\dots} &\to &M^{a\dots}_{\phantom{a\dots}b\dots}.
\end{eqnarray}
It is easy to verify that the flatness of $\gamma^{ab}$ is maintained by this map, as $\breve{R}^a_{\phantom{a}bcd} \to (\phi^*\breve{R})^a_{\phantom{a}bcd}$ and $\phi^* 0 = 0$.\footnote{These transformations form a subgroup of a larger group of transformations for which $\gamma^{ab}\to \gamma'^{ab}$ (still flat) and $h^{ab}$ compensates such that $g^{ab}$ is held fixed. Because this larger group does not relate so simply to the diffeomorphism gauge freedom, it is not discussed here.} It should be noted that the replacement $\gamma^{ab}\to(\phi^*\gamma)^{ab}$ does not represent a coordinate change, but is a map between two different metric tensors. Obviously, because both metrics are flat, we can always find coordinates for each such that their components are those of the Minkowski matrix $\eta^{\alpha\beta}=\mathrm{diag}(+1,-1,-1,-1)$, but while $\gamma^{\alpha\beta}=\eta^{\alpha\beta}$ in some coordinates $\{x^\alpha\}$, in general $(\phi^*\gamma)^{\alpha\beta}=\eta^{\alpha\beta}$ in a different set of coordinates $\{y^\alpha\}$.

A key feature of the $\gamma$-transformation (\ref{gen}) is that it allows us to distinguish between the two types of tensors in Grishchuk's formulation: those that can be constructed in standard GR, and `non-GR' tensors, which cannot. Because $g^{ab}$ is invariant under (\ref{gen}), all GR tensors (which must be expressible in terms of $g^{ab}$, $\nabla_a$ and $M^{a\dots}_{\phantom{a\dots}b\dots}$ only) will be likewise unchanged:
\begin{eqnarray}
\textrm{GR:} \quad A^{a\dots}_{\phantom{a\dots}b\dots}\to A^{a\dots}_{\phantom{a\dots}b\dots}.
\end{eqnarray}
Thus, any tensor which is not invariant under \emph{all} transformations of the form (\ref{gen}) must be non-GR:
\begin{eqnarray}
\textrm{non-GR:} \quad B^{a\dots}_{\phantom{a\dots}b\dots}\to {B'}^{a\dots}_{\phantom{a\dots}b\dots} \ne B^{a\dots}_{\phantom{a\dots}b\dots},
\end{eqnarray}
for some $\gamma$-transformation. 

From this identification, and the formula (\ref{tab}), we can confirm our suspicions that $t^{ab}$ is a non-GR quantity: under a $\gamma$-transformation (\ref{gen}), the $g$'s in the square brackets are untouched, but the $\gamsem$ derivatives are transformed according to
\begin{eqnarray}\label{gamgra-trans}
\gamgra_a &\to& {\gamgra'}_a, \nonumber\\
\textrm{where}\quad\qquad \gamgra_a\gamma^{bc}&=& 0,\nonumber\\
 \textrm{and}\qquad  {\gamgra'}_a(\phi^*\gamma)^{bc}&=& 0.
\end{eqnarray}
Although there may be some $\phi$ for which the transformation of $1/\gamma$ in (\ref{tab}) cancels the effects of transformation of $\gamgra_a$, this  will not happen for all $\phi$.\footnote{To demonstrate this rigorously it is sufficient to show that $t^{ab}$ is not invariant under infinitesimal $\gamma$-transformations; this calculation is performed in the appendix.} Thus $t^{ab}$ is not in general an invariant of the transformation, and must be impossible to construct in GR without introducing additional structure in the form of $\gamma^{ab}$. 

Clearly, it is important to know whether the $\gamma$-transformation should be thought of as a gauge transformation, or as map between physically inequivalent systems. This is not a trivial problem, however, because we must be careful to avoid the tacit assumption that the GR metric $g^{ab}$ describes everything about the gravitational field. Because $g^{ab}$ is invariant under (\ref{gen}), the physics traditionally thought of as spacetime `geometry' (and, in the flat-space view, are the observable effects of $h^{ab}$ on particle worldlines, rods and clocks) must be left invariant also. Thus, comparing the $\gamma$-dependence of $t^{ab}$ with the $\gamma$-independence of spacetime `geometry'\footnote{We insist on writing `geometry' in inverted commas because although the phenomena to which we are referring are traditionally deemed to be the result of spacetime geometry, we must stress that this interpretation is not endorsed by Grishchuk's formulation. The term `geometry' in this sense should simply be taken as a short-hand for the observable predictions shared by General Relativity and the flat-space formalism.}, we can immediately conclude that that $t^{ab}$ cannot be determined by spacetime `geometry' alone. However, it does not immediately follow that $t^{ab}$ is an unphysical tensor, as we must seriously examine the possibility that gravity is more than just $g^{ab}$, and that in performing the $\gamma$-transformation we have altered something physical about the system that standard General Relativity simply does not `see'.

If we suppose that (\ref{gen}) does effect a physically meaningful change, we must conclude that every physical system is associated with a `true' $\gamma^{ab}$, or at least with a class of physically equivalent flat-metrics $\{\gamma^{ab}\}$ that is smaller than the complete space spanned by all possible $\gamma$-transformations. The question is, given a physical system, how can we know when we have chosen the correct $\gamma^{ab}$? Clearly, no `geometric' measurements can ever reveal which $\gamma^{ab}$ is hidden beneath the $g^{ab}$ metric, because `geometric' phenomena are invariant under the $\gamma$-transformation. The only possibility of revealing $\gamma^{ab}$ empirically would be if we could directly measure a non-GR tensor like $t^{ab}$. However, to assume that such a measurement could be carried out would make our logic circular, as for that to be possible the tensor would certainly need to be \emph{physically meaningful}, and it is the truth of precisely this assertion that we have been trying to determine!

Even if we cannot rely on an empirical method to reveal the `true' flat-metric $\gamma^{ab}$ of a particular physical system, there may still be a systematic way to \emph{define} one, given knowledge of quantities we can measure. Such a definition would pick out a `canonical' $\gamma^{ab}$ and we would be forbidden from performing $\gamma$-transformations because the new $\gamma^{ab}$ would no longer be canonical.\footnote{One might expect $\gamma^{\alpha\beta}=\eta^{\alpha\beta}$ to be a perfectly good definition for a canonical flat-metric; however this does not really fix $\gamma^{ab}$ at all, it only begs the question: in which coordinate system do we insist that this equation holds?} The situation is analogous to the following question in electrostatics: what is the potential $V$ at a particular point $x$? Even though we can never measure this quantity directly, we can still define a canonical potential $V(x)$ in a natural and systematic way by demanding that $V\to0$ as the distance from the sources $r\to \infty$, or equivalently, as the electric charges  $q_i \to0$. In the same sense that we have $V=0$ (everywhere) synonymous with the absence of electric charges, we would certainly hope that we could choose a canonical $\gamma^{ab}$ such that $h^{ab}=0$ (everywhere) is synonymous with the absence of matter fields. Indeed, given a GR metric $g^{ab}$ that satisfies the Einstein field equations with a matter energy-momentum tensor $T^{ab}$ as the source, we can write:
\begin{eqnarray}\label{sol}
g^{ab} = g^{ab}(T^{cd}),
\end{eqnarray}
and define the canonical flat metric by
\begin{eqnarray}\label{nat}
\gamma^{ab} = g^{ab}(T^{cd})|_{T^{cd}=0\ \!\textrm{(everywhere)}}.
\end{eqnarray}
For example, we could view the Schwarzschild spacetime with central mass $M$ as a family of spacetimes $g^{ab}(M)$ and identify $\gamma^{ab}$ with $g^{ab}(0)$. For any other prescription for the canonical $\gamma^{ab}$ there will arise the following peculiar situation: in the absence of matter, despite spacetime `geometry' being flat, $g^{ab}$ will be not be equal to $\gamma^{ab}$, and we will still have to use a non-zero $h^{ab}$ field to convert between these two different flat metrics. In this sense (\ref{nat}) is the only natural prescription for a canonical flat metric.

However, it turns out that even this effort cannot force us to abandon (\ref{gen}) as a genuine gauge transformation, as (\ref{nat}) does not behave correctly under some diffeomorphism gauge transformations (DGTs). To see this, start with a GR metric $g^{ab} =  g^{ab}(T^{cd})$ and a canonical flat metric defined by (\ref{nat}).  Now, consider a family of diffeomorphisms $\{\phi_f : \mathcal{M} \to \mathcal{M} \quad \forall  f \in \mathbb{R}\}$ such that $\phi_0$ is the identity diffeomorphism: $\phi_0(p)=p \ \ \forall p\in \mathcal{M}$. We change nothing physical about this system by performing a DGT with $\phi_{f}$ for any value of the parameter $f$, and we are free to have the value of $f$ determined by some functional of $T^{cd}$ such that $T^{cd} = 0 \ \textrm{(everywhere)}$ gives $f =0$. Then, having performed this DGT, we can calculate the canonical flat metric again:
\begin{eqnarray}
{\gamma'}^{ab} &=& {g'}^{ab}({T'}^{cd})|_{{T'}^{cd}=0} = \left[(\phi_f^*g)^{ab}({T'}^{cd})\right]_{{T'}^{cd}=0}\nonumber\\
&=& (\phi_0^*\gamma)^{ab} = \gamma^{ab}.
\end{eqnarray}
Thus, our DGT, coupled with our definition of the natural canonical flat metric, has had the following effect:
\begin{eqnarray}\label{fixed}
g^{ab} &\to& (\phi_f^*g)^{ab},\nonumber \\
\gamma^{ab}&\to& \gamma^{ab},  \nonumber\\
h^{ab} &\to& {h'}^{ab} = \tfrac{\sqrt{-\phi_f^*\gamma}}{\sqrt{-\gamma}}\left( (\phi_f^*\gamma)^{ab} + (\phi_f^*h)^{ab} \right) - \gamma^{ab}, \nonumber \\
M^{a\dots}_{\phantom{a\dots}b\dots} &\to &(\phi_f^*M)^{a\dots}_{\phantom{a\dots}b\dots}.
\end{eqnarray}
Whereas, under the DGT, we should have recovered $\gamma^{ab} \to (\phi_f^*\gamma)^{ab}$ and $h^{ab} \to (\phi_f^*h)^{ab}$. We are left with a choice: either we completely abandon the idea of a natural canonical $\gamma^{ab}$ on the grounds that it is not covariant under all DGTs (and thus accept that the $\gamma$-transformation (\ref{gen}) \emph{is} a gauge transformation), or we agree that this `$\gamma$-fixed' transformation (\ref{fixed}) is on equal footing with a DGT and is therefore another gauge transformation of the formalism. Of course, this is not really a choice at all, as the $\gamma$-fixed transformation has precisely the same effect as performing a diffeomorphism gauge transformation with $\phi_f$ and then a $\gamma$-transformation with $(\phi_f)^{-1}$; thus, by agreeing that (\ref{fixed}) is a gauge transformation, one has agreed that the $\gamma$-transformation is one also. 

The key to this argument is that because the prescription (\ref{nat}) does not pick $\gamma^{ab}$ in a diffeomorphism covariant fashion\footnote{We implicitly picked a gauge when we wrote $g^{ab}$ as a \emph{particular} solution of the field equations with source $T^{cd}$ in (\ref{sol}).}, we retain the ability to perform $\gamma$-transformations through our choice of which diffeomorphism gauge we use to express the $T^{ab}=0$ spacetime when we apply the definition for the canonical flat metric.

It is interesting to note that when Grishchuk refers to the gauge transformations of his formalism in \cite{G}, he appears to mean the $\gamma$-fixed variety: in the appendix we calculate that the effect of an infinitesimal $\gamma$-fixed transformation on $h^{ab}$ is
\begin{eqnarray}\label{h trans}
h^{ab} \to {h'}^{ab} &=& h^{ab} + \left(\xi^{c}(\gamma^{ab} +h^{ab})\right)_{\gamsem c} \nonumber\\
&&{} - 2 \xi^{(a}_{\phantom{(a}\gamsem c}\left(\gamma^{b)c}+ h^{b)c}\right)
\end{eqnarray}
and on setting $\gamma^{\alpha\beta}=\eta^{\alpha\beta}$ (which can either be viewed as a coordinate choice, given $\gamma^{ab}$, or a choice of $\gamma^{ab}$ given some coordinate system) we recover 
\begin{eqnarray}
{h'}^{\alpha \beta} &=& h^{\alpha\beta} + \eta^{\alpha\beta}\xi^\lambda_{\phantom{\lambda}, \lambda} + (h^{\alpha\beta}\xi^\lambda)_{,\lambda}\nonumber \\ && {} - 2\xi^{(\alpha,\beta)} - 2\xi^{(\alpha}_{\phantom{(\alpha},\lambda}h^{\beta) \lambda}
\end{eqnarray}
which is equation (38) of \cite{G}.

Thus we must finally conclude that the $\gamma$-transformation (\ref{gen}) \emph{is} a gauge transformation of Grishchuk's formalism, and that not only is the flat metric $\gamma^{ab}$ unobservable, it is impossible to define a `canonical' choice of $\gamma^{ab}$ in a diffeomorphism gauge covariant, systematic, and natural fashion.

\subsection{The transformation properties of $t^{ab}$}\label{transtab}
We have demonstrated that the $\gamma$-transformation should be thought of as a map between different mathematical representations of the same physical system. As $t^{ab}$ is not invariant under this gauge change (i.e.~non-GR) we might be suspicious that this `energy-momentum tensor' has no physical significance. However, before we dismiss $t^{ab}$, it is worth considering the following possibility: even though $t^{ab}$ is not invariant under $\gamma$-transformations, could the transformed tensor $t'^{ab}$, somehow, have the same \emph{physical content} as the untransformed tensor $t^{ab}$? After all, we see exactly this behaviour for a DGT: no tensor field is \emph{invariant} under (\ref{dif}), however we can consider tensor fields to be \emph{covariant} under this transformation (and their physical content unaltered) because they allow for the construction of gauge invariant quantities.\footnote{All measurements necessarily correspond to scalars, thus the action of a DGT is simply to move these scalars to different points of $\mathcal{M}$. Because all the worldlines of observers and test particles are similarly displaced, the correlations between these scalars will be diffeomorphism gauge invariant.} We must therefore consider the possibility that the $\gamma$-transformation law for $t^{ab}$ constitutes some form of `generalised covariance' that would allow gauge invariant quantities to be constructed.

Of course, the expected form of these invariants rather depends on what one supposes the physical content of $t^{ab}$ to be. If it is, indeed, an energy momentum tensor, then an observer with 4-velocity $u^a$ would expect to `find' some energy density $\rho=t^{ab}u^c u^d g_{ac}g_{bd}$, or possibly $\rho=t^{ab}u^c u^d \gamma_{ac}\gamma_{bd}$. It is easy to check that neither of these quantities are invariant under a $\gamma$-transformation, despite the fact that we were forced to conclude that these transformations do not alter whatsoever the physical system we are examining. From this we deduce that, whatever physical meaning $t^{ab}$ may have, since it cannot define a meaningful energy-density in the standard way, it is definitely not an energy-momentum tensor. 

\subsubsection{Infinitesimal transformations}
It is instructive to examine the transformation properties of $t^{ab}$ for an arbitrary infinitesimal gauge transformation. We proceed by constructing a diffeomorphism very close to the identity by Lie dragging tensor fields along an infinitesimal vector field $\xi^{a}$:
\begin{eqnarray}\label{lie}
(\phi^*X)^{a\dots}_{\phantom{a\dots}b\dots} = X^{a\dots}_{\phantom{a\dots}b\dots} + (\mathcal{L}_\xi X)^{a\dots}_{\phantom{a\dots}b\dots},
\end{eqnarray}
where $\mathcal{L}_\xi$ is the Lie derivative along $\xi^{a}$. Under a $\gamma$-fixed gauge transformation for an infinitesimal diffeomorphism $\phi$ defined by (\ref{lie}), we find that $t^{ab} \to t'^{ab}$, where
\begin{eqnarray}\label{tabtrans}
\kappa {t'}^{ab} &=& \kappa \left(t^{ab} +  (\mathcal{L}_\xi t)^{ab}\right) \nonumber \\
&& {} + \left[\xi^{e}_{\phantom{e}\gamsem e}\left(\frac{g}{\gamma}\left(g^{ab}g^{cd}-g^{a(c}g^{d)b}\right)\right)_{\gamsem c}\right]_{\gamsem d}\nonumber \\
&& {} - \xi^{e}_{\phantom{e}\gamsem c \gamsem d}\left(\frac{g}{2\gamma}\left(g^{ab}g^{cd}-g^{ac}g^{db}\right)\right)_{\gamsem e}.
\end{eqnarray}
This result is calculated in the appendix. An important point of (\ref{tabtrans}) is that, unlike the $\gamma$-fixed behaviour of a GR field ($A\to A + \mathcal{L}_\xi A$), the transformation law for $t^{ab}$ includes second derivatives of $\xi$. Thus, in a qualitative sense, the new $t'^{ab}$ (evaluated at some point $p\in \mathcal{M}$) seems to depends much more on the details of the transformation than a GR quantity would; certainly the complex formula (\ref{tabtrans}) cannot be interpreted as some simple algebraic or geometric operation.   If we imagine producing a finite transformation by `exponentiating' (\ref{tabtrans}) then the GR part of the transformation $t^{ab} + (\mathcal{L}_\xi t)^{ab}$ would correspond (loosely speaking) to a diffeomorphism ` $\phi^* = \textrm{e}^{\mathcal{L}_\xi}$ ' which would, to first order in $\xi$, only depend on $\xi$ and its first derivatives. The extra terms in (\ref{tabtrans}), once exponentiated, would vastly increase our freedom to determine $t'^{ab}$ at any particular $p$, possibly enough to set $t'^{ab}(p)=0$ for any $t^{ab}$. If this were indeed shown to be the case, then $t^{ab}$ could hardly represent a meaningful \emph{local} property of any field.

A particularly undesirable feature of (\ref{tabtrans}) is that $t'^{ab}$ is not determined by $\xi^a$ and $t^{ab}$ alone; we also need to know the tensor $[(g / 2 \gamma)(g^{ab}g^{cd}-g^{ac}g^{db})]_{\gamsem e}$ from which $t^{ab}$ has been constructed. This detail seems to preclude the assembly of invariants from $t^{ab}$ and observer worldlines alone.\footnote{Because non-GR tensors can be combined to form GR tensors, it will always be possible to `add in' some combination of $\gamma^{ab}$, $h^{ab}$, and $\gamgra_a$ to create a gauge invariant quantity from $t^{ab}$. However, in this case we should not associate the invariants with $t^{ab}$ by itself, but instead with the larger GR object we have assembled.}

\subsubsection{Finite transformations}\label{finite}
To study the effect of finite gauge transformations on $t^{ab}$, we focus on the Schwarzschild spacetime with a central point-mass $M$. Working in natural units $(c=G=1)$ and suppressing the abstract indices on the coordinate differentials $(\ud x^\alpha)_a$, we write the GR metric as
\begin{eqnarray}\label{fg}
g_{\alpha\beta}\ud x^{\alpha}\ud x^{\beta} &=& \frac{1}{(f_1 g_1 - f_2 g_2)^2}\Big[(g_1^2 - g_2^2)\ud t^2 \nonumber \\ 
&& {} + 2(f_1 g_2 - f_2 g_1)\ud t \ud r - (f_1^2 - f_2^2) \ud r^2\Big]  \nonumber \\
&& {} - r^2 (\ud \theta^2 + \mathrm{sin}^2\theta \ud \phi^2),
\end{eqnarray}
where $\{f_1, f_2, g_1, g_2\}$ are functions of $r$ and $t$ only. Birkhoff's theorem \cite{Birk} shows the Schwarzschild spacetime to be the \emph{only} spherically symmetric vacuum solution to the Einstein equations; thus for any choice of $\{f_i, g_i\}$ consistent with $R^{ab}=0$, the metric given by (\ref{fg}) represents the Schwarzschild spacetime. This form of $g^{ab}$ will be particularly useful for the present discussion, as it will allow us to choose explicitly the `gauge' in which to express the gravitational field. To illustrate this point, we record below the recipes for the commonly used representations of the Schwarzschild spacetime. 
\begin{displaymath}
\begin{array}{r|ccc}
 & \begin{array}{c} $Standard$\\$Schwarzschild$\end{array} & \begin{array}{c} $Advanced$\\$Eddington-$ \\  $Finkelstein$ \end{array} & \begin{array}{c} $Painlev\'e-$\\$Gullstrand$ \end{array} \\
\hline
f_1 & 1/\sqrt{1 - 2M/r} & 1 +M/r & 1 \\
f_2 & 0 & M/r & 0 \\
g_1 & \sqrt{1 - 2M/r} & 1-M/r & 1 \\
g_2 & 0&-M/r & -\sqrt{2M/r} \\
\end{array}
\end{displaymath}

As we have emphasised, there is no unique $\gamma^{ab}$ hidden beneath the metric defined in (\ref{fg}). However, for the sake of concreteness, we fix the flat-metric as 
\begin{eqnarray}\label{gammachoice}
\gamma_{\alpha\beta}\ud x^{\alpha}\ud x^{\beta} &=&\ud t^2 - \ud r^2  - r^2 (\ud \theta^2 + \mathrm{sin}^2\theta \ud \phi^2),
\end{eqnarray}
so that altering the functions $\{f_i, g_i\}$ will give rise to $\gamma$-fixed transformations.\footnote{Equally we could have arranged for this process to run in the opposite direction. Starting with the standard form of the Schwarzschild metric in $(t,r, \theta, \phi )$ coordinates, we could have  performed a coordinate transformation to a system $(T,R, \theta, \phi )$ that preserved the spherical symmetry. Working in these coordinates, a seeming natural choice of the flat-metric would have been $\gamma_{\alpha\beta}\ud x'^{\alpha}\ud x'^{\beta}=\ud T^2 - \ud R^2  - R^2 (\ud \theta^2 + \mathrm{sin}^2\theta \ud \phi^2)$, a different tensor from the one defined by (\ref{gammachoice}). The choice of coordinates used to represent $g^{ab}$ would therefore determine $\gamma^{ab}$ but not alter $g^{ab}$ itself. The effect would be that of a $\gamma$-transformation.}

In order to proceed, we remove some of the gauge freedom by demanding
\begin{eqnarray}\label{cond}
\begin{array}{lrcl}
1. &\quad f_2 &=& 0, \\
2. &\quad \partial_t g_{\alpha\beta} &=& 0.
\end{array}
\end{eqnarray} 
Then we find that the vacuum field equations $R^{ab}=0$ enforce
\begin{eqnarray}\label{vac1}
f_1 g_1 &=& C,\\ \label{vac2} 
g_1^2 - g_2^2 &=& 1 - 2M/r,
\end{eqnarray}
where $C$ and $M$ are constants, and we have identified the latter as the central mass by comparison with the Standard Schwarzschild and Painlev\'e-Gullstrand gauges. Under these conditions we find that all the components of $t^{ab}$ vanish apart from the energy density:
\begin{eqnarray}
t^{\alpha i} &= &t^{ i \alpha} = 0, \\ \label{gauges}
t^{00}&= &- \frac{g_1^3 - g_1 + 2r\partial_r g_1}{g_1^3 r^2}.
\end{eqnarray}
This last formula makes manifest the large space of gauge equivalent energy-momentum tensors associated with the Schwarzschild spacetime, even after we have removed a large portion of gauge freedom by demanding (\ref{cond}). Notice in particular that the Standard Schwarzschild gauge yields
\begin{eqnarray}\label{SW}
t^{00}&= &- \left(\frac{2M}{r(r-2M)}\right)^2,
\end{eqnarray}
whereas, in the Painlev\'e-Gullstrand gauge
\begin{eqnarray}
t^{00}=0.
\end{eqnarray}
The gauge equivalence of these two results leaves little room for a physical interpretation of this energy-momentum tensor. Because $t^{ab}$ can be made to vanish everywhere by a gauge transformation, it cannot possibly convey any more gauge-invariant information than to tell us that this spacetime is empty of whatever it is that $t^{ab}$ represents. While this is not unreasonable per se (as $t^{ab}$ might only be sensitive to gravitational radiation or some other phenomena absent from the Schwarzschild spacetime) it then becomes very difficult to justify why the energy-momentum tensor should be non-zero in any gauge at all. This uncomfortable situation would force us to identify a whole host of non-trivial energy-momentum tensors with emptiness, of which (\ref{gauges}) are only a small fraction.

As the Advanced Eddington-Finkelstein gauge has $f_2\ne0$, we cannot use (\ref{gauges}) to calculate the energy-momentum tensor. Instead, we take the general formula (\ref{tab}) as our starting point, and recover $t^{ab}=0$, just as we found in the Painlev\'e-Gullstrand gauge. This agreement suggests that the non-zero energy-momentum tensor (\ref{SW}) might only be an artefact of the `horizon' present in the Standard Schwarzschild gauge: in the $(t,r,\theta, \phi)$ coordinate system picked out by $\gamma_{ab}$, the components of the GR metric $g_{\alpha\beta}$ are singular at $r=2M$. In contrast, Painlev\'e-Gullstrand and Advanced Eddington-Finkelstein are \emph{global} gauges: the components $g_{\alpha\beta}$ are regular everywhere but at the origin. While a coordinate singularity is admissible within differential geometry, in Grishchuk's flat-space picture this would correspond to an infinite `gravitational field' $h^{ab}$, which could be deemed unphysical. This line of reasoning allows us to reject (\ref{SW}) because it was derived in a gauge which transforms the gravitational field to infinity at some points, and we would then hope to confirm that the physical result ($t^{ab}=0$) applies in all global gauges. Unfortunately, this turn out to be impossible, as we show by means of a counter-example. Consider a family of gauges parametrised by $\lambda$:
\begin{eqnarray}
\begin{array}{rcl}
f_1 &=& \sqrt{r/(r + \lambda M)}, \\
f_2 &=& 0, \\
g_1 &=& \sqrt{(r + \lambda M)/{r}}, \\
g_2 &=& \sqrt{(2 + \lambda)M/r}.
\end{array}
\end{eqnarray} 
It is easy to check that these obey the restrictions (\ref{cond}) and the vacuum field equations (\ref{vac1}) and (\ref{vac2}). Furthermore, for $\lambda \ge 0$, $g_{\alpha\beta}$ defined by (\ref{fg}) is regular everywhere but the origin. Using (\ref{gauges}) we find that (apart from $\lambda=0$ which is just Painlev\'e-Gullstrand again) the energy-momentum tensor is non-zero:
\begin{eqnarray}
t^{00}= - \left(\frac{\lambda M}{r(r + \lambda M)}\right)^2.
\end{eqnarray}
Not only can we make $ t^{00}(r)$ take on a wide range of values by adjusting $\lambda$, we also note that in the limit  $\lambda \to \infty$, we have the disconcerting situation of a non-zero energy density that is independent of $M$.

In light of all these results, it appears highly unlikely that the behaviour of $t^{ab}$ would permit the extraction of gauge invariant information and allow us to view this tensor as maintaining some physical content under gauge transformations.

\section{Conclusion}
The formulation of gravity presented in \cite{G} succeeds in recasting General Relativity as a flat-space theory of a symmetric tensor field. While we do not find fault with the formalism itself, we assert that care must by taken in its interpretation, as we believe we have demonstrated that only those quantities which can be defined solely in terms of GR tensors are of any physical importance. The physically insignificant content of the flat-space formalism is a consequence of an unmeasurable field $\gamma^{ab}$ which is not uniquely determined by the requirement that it be a flat metric tensor.

There are in principle two ways to deal with the non-uniqueness of $\gamma^{ab}$: 1.~Pick a particular flat metric and declare that this is the immutable `correct' choice, to be used in all situations; 2.~Allow $\gamma^{ab}$ to depend somehow on the physical system we are describing, or how we have chosen to represent the system mathematically. 

The problem with the first stance is that the theory still retains $\gamma$-fixed gauge transformations. To see this, note that equation (53) of \cite{BG} expresses the equivalence of Grishchuck's equations of motion ($r_{ab} \equiv - P^c_{\phantom{c}ab\gamsem c} - P^c_{\phantom{c}ad}P^d_{\phantom{d}bc} + \tfrac{1}{3}P_a P_b =0$) with the Einstein field equations:
\begin{eqnarray}
R_{ab}[g]=\breve{R}_{ab}[\gamma] + r_{ab}[h,\gamma].
\end{eqnarray}
Babak and Grishchuk interpret this relation as follows: given a flat-metric $\gamma^{ab}$, an $h^{ab}$ that satisfies $r_{ab}=0$ will enforce $R_{ab}=0$, establishing the agreement with GR. However, one can always use this equation to make the converse argument: given a flat-metric $\gamma^{ab}$, a $g^{ab}$ which solves $R_{ab}=0$ will enforce Grishchuk's equation $r_{ab}=0$. As $R_{ab}[\phi^* g]=0$ if $R_{ab}[g]=0$, we can construct a whole range of solutions $\{h^{\prime ab}: r_{ab}[h^\prime]=0\}$ from $h^{ab}$ simply by applying diffeomorphisms to $g^{ab}$. Because we declared $\gamma^{ab}$ to be immutable, these new solutions will correspond to $\gamma$-fixed transformations of $h^{ab}$. Crucially, as $g^{\prime ab}= \phi^*g^{ab}$, no `geometric' experiment can tell any $h^{\prime ab}$ apart from from $h^{ab}$. Thus, without a method to measure a non-GR quantity directly, we have to conclude that these new solutions represent physically equivalent systems, and that the $\gamma$-fixed transformation is a gauge transformation of the theory.

The second stance appears to be able to dodge this argument, because one can claim that we should have applied the same diffeomorphism to $\gamma^{ab}$ that we applied to $g^{ab}$, forcing us to perform a harmless DGT instead of a $\gamma$-fixed transformation. However, if we take this view, we will need a heuristic for deriving $\gamma^{ab}$ from measurable quantities, otherwise we will never know where to start with the `correct' pairing $(g^{ab}, \gamma^{ab})$. In order that this heuristic be consistent with arbitrary DGTs (which are gauge transformations of any tensorial theory) any prescription for which  $T^{ab}=0 \Rightarrow h^{ab}=0$ inevitably leads us to identify $\gamma$-transformations as gauge transformations anyway, because we are free to represent the $T^{ab}=0$ limit in any diffeomorphism gauge we choose.

Accepting that $\gamma$-transformations and $\gamma$-fixed transformation are maps between different mathematical representations of the same physical system, we conclude that the exotic gauge transformation properties of $t^{ab}$ cannot allow us to interpret this tensor as a local measure of the energy and momentum content of the gravitational field. Although $t^{ab}$ is a perfectly legitimate mathematical construction, its dependence on the unmeasurable and non-unique tensor $\gamma^{ab}$ renders it ill-defined, and devoid of physical meaning.

\section{Acknowledgements}
We are grateful to L.~P.~Grishchuk for his valuable comments, although we should make it clear that he disagrees with our conclusions. LMB thanks STFC for their support. ANL thanks Michael Ibison for helpful initial discussions, and for provision of a program for calculating the elements of $t^{\alpha\beta}$ in the standard Schwarzschild gauge. (This used equation (65) of \cite{BG}, and thus allowed an independent check of the standard  Schwarzschild gauge results discussed in Section \ref{finite}, which were calculated using (\ref{tab}).)

\section*{Appendix: Infinitesimal transformations}
Here we calculate how $h^{ab}$, $t^{ab}$, and $\gamgra_a$ change under transformations defined by diffeomorphisms infinitely close to the identity: $\phi^* = 1 + \mathcal{L}_\xi$. In this limit, the $\gamma$-fixed transformation (\ref{fixed}) for $h^{ab}$ is
\begin{eqnarray}
h^{ab}&\to& {h'}^{ab} \\
{h'}^{ab}&=& (-\gamma)^{-1/2} \left(1 + \mathcal{L}_\xi \right)\left(\sqrt{-\gamma}(\gamma^{ab} + h^{ab})\right) - \gamma^{ab}\nonumber\\
&=& (-\gamma)^{-1/2}\mathcal{L}_\xi \left(\sqrt{-\gamma}(\gamma^{ab} + h^{ab})\right) + h^{ab}.\nonumber
\end{eqnarray}
Thus,
\begin{eqnarray}
\delta h^{ab} &\equiv& {h'}^{ab} - h^{ab} \nonumber\\
&=& (\gamma^{ab} + h^{ab})(-\gamma)^{-1/2}\mathcal{L}_\xi \sqrt{-\gamma} + \mathcal{L}_\xi (\gamma^{ab} + h^{ab})\nonumber\\
&=& (\gamma^{ab} + h^{ab})\xi^{c}_{\phantom{c}\gamsem c} + \xi^{c}\left(\gamma^{ab} + h^{ab}\right)_{\gamsem c} \nonumber\\
&& {} - 2 \xi^{(a}_{\phantom{(a}\gamsem c}\left(\gamma^{b)c}+ h^{b)c}\right) \nonumber\\
&=& \left(\xi^{c}(\gamma^{ab} +h^{ab})\right)_{\gamsem c} - 2 \xi^{(a}_{\phantom{(a}\gamsem c}\left(\gamma^{b)c}+ h^{b)c}\right),
\end{eqnarray}
proving (\ref{h trans}).

To calculate the behaviour of the energy-momentum tensor under a $\gamma$-fixed transformation, we define the tensor
\begin{eqnarray}\label{Y}
Y^{abcd}\equiv \frac{g}{\gamma}g^{(a[b)}g^{(c]d)} = \frac{g}{2\gamma}\left(g^{ab}g^{cd}-g^{a(c}g^{d)b}\right),
\end{eqnarray}
so that
\begin{eqnarray}
\kappa t^{ab} = Y^{abcd}_{\phantom{abcd}\gamsem c \gamsem d}.
\end{eqnarray}
Under the $\gamma$-fixed transformation, $t^{ab} \to t'^{ab}$ where
\begin{eqnarray}\label{tabtrans1}
\kappa {t'}^{ab} &=& \left[\gamma^{-1}(1+\mathcal{L}_\xi)\left(gg^{(a[b)}g^{(c]d)}\right)\right]_{\gamsem c \gamsem d}\nonumber\\
&=& \kappa t^{ab} + \left[\mathcal{L}_\xi\left(Y^{abcd}\right) - gg^{(a[b)}g^{(c]d)}\mathcal{L}_\xi(\gamma^{-1}) \right]_{\gamsem c \gamsem d}\nonumber\\
&=& \kappa t^{ab} + \Big[Y^{abcd}_{\phantom{abcd}\gamsem e} \xi^{e} - 2 \xi^{(a}_{\phantom{(a}\gamsem e}Y^{b)ecd}\nonumber\\
&& \qquad \qquad {} - 2 Y^{abe(c}\xi^{d)}_{\phantom{d)}\gamsem e} +2 Y^{abcd}\xi^{e}_{\phantom{e}\gamsem e}\Big]_{\gamsem c \gamsem d}.\nonumber
\end{eqnarray}
In contrast, were $t^{ab}$ a GR tensor, under the $\gamma$-fixed transformation we would have $t^{ab} \to t^{ab} + \mathcal{L}_\xi t^{ab}$, with
\begin{eqnarray}
\kappa\mathcal{L}_\xi t^{ab}= \xi^{e}Y^{abcd}_{\phantom{abcd}\gamsem c\gamsem d\gamsem e} - 2\xi^{(a}_{\phantom{(a}\gamsem e}Y^{b)ecd}_{\phantom{b)ecd}\gamsem c \gamsem d}.\nonumber
\end{eqnarray}
Thus, the non-GR part of $\kappa t'^{ab}$ is
\begin{eqnarray}
\Delta (\kappa t^{ab})&\equiv&\kappa \left({t'}^{ab} -  t^{ab} - \mathcal{L}_\xi t^{ab}\right)\nonumber\\
&=& \xi^{e}_{\phantom{e}\gamsem c \gamsem d}Y^{abcd}_{\phantom{abcd}\gamsem e} + 2 \xi^{e}_{\phantom{e}\gamsem c}Y^{abcd}_{\phantom{abcd} \gamsem e \gamsem d}\nonumber\\
&& {} - 2\xi^{(a}_{\phantom{(a}\gamsem e \gamsem c \gamsem d}Y^{b)ecd} - 4\xi^{(a}_{\phantom{(a}\gamsem e \gamsem c}Y^{b)ecd}_{\phantom{b)ecd}\gamsem d}\nonumber\\
&& {} - 2\left[Y^{abe(c}\xi^{d)}_{\phantom{d)}\gamsem e} - Y^{abcd}\xi^{e}_{\phantom{e}\gamsem e}\right]_{\gamsem c \gamsem d}.
\end{eqnarray}
Note that the third and fourth terms vanish because $Y^{abcd}=-Y^{acbd}$ and $\gamgra_a$ operators commute. Expanding out the derivatives acting on the square brackets, then cancelling and collecting like terms, we arrive at
\begin{eqnarray}
\Delta (\kappa t^{ab})&=& 2\xi^{e}_{\phantom{e}\gamsem e}Y^{abcd}_{\phantom{abcd}\gamsem c \gamsem d} - \xi^{e}_{\phantom{e}\gamsem c \gamsem d}Y^{abcd}_{\phantom{abcd}\gamsem e} + 2\xi^{e}_{\phantom{e}\gamsem e \gamsem d}Y^{abcd}_{\phantom{abcd}\gamsem c}\nonumber \\
&=& 2 \left[\xi^{e}_{\phantom{e}\gamsem e}Y^{abcd}_{\phantom{abcd} \gamsem c}\right]_{\gamsem d} - \xi^{e}_{\phantom{e}\gamsem c \gamsem d}Y^{abcd}_{\phantom{abcd}\gamsem e}.
\end{eqnarray}
Replacing $Y^{abcd}$ with its definition (\ref{Y}), the transformation law (\ref{tabtrans}) immediately follows. 

Because the $\gamma$-fixed transformation is simply a DGT with $\phi$ followed by a $\gamma$-transformation with $\phi^{-1}$, it is easy to use this result to calculate the behaviour of $t^{ab}$ under an infinitesimal $\gamma$-transformation:
\begin{eqnarray}
(\gamma\textrm{-fixed})_{\phi}t^{ab} &=& \left(\gamma\textrm{-trans}\right)_{\phi^{-1}}\left(\textrm{DGT}\right)_{\phi}t^{ab} \nonumber \\
&=& \left(\gamma\textrm{-trans}\right)_{\phi^{-1}}\left(t^{ab} + \mathcal{L}_\xi t^{ab}\right)\nonumber \\
&=& \left(\gamma\textrm{-trans}\right)_{\phi^{-1}} t^{ab} + \mathcal{L}_\xi t^{ab}, 
\end{eqnarray}
for infinitesimal $\xi$. Thus, under a $\gamma$-transformation, $\kappa t^{ab}$ becomes 
\begin{eqnarray}
\kappa {t'}^{ab} &=& \kappa t^{ab} - \left[\xi^{e}_{\phantom{e}\gamsem e}\left(\frac{g}{\gamma}\left(g^{ab}g^{cd}-g^{a(c}g^{d)b}\right)\right)_{\gamsem c}\right]_{\gamsem d}\nonumber \\
&& {} + \xi^{e}_{\phantom{e}\gamsem c \gamsem d}\left(\frac{g}{2\gamma}\left(g^{ab}g^{cd}-g^{ac}g^{db}\right)\right)_{\gamsem e},
\end{eqnarray}
which clearly demonstrates that $t^{ab}$ is non-GR. 

For completeness, we calculate how the derivative operator $\gamgra_a$ changes under an infinitesimal $\gamma$-transformation. We shall proceed without using the flatness of $\gamma^{ab}$, in order that the result be in its most general form; only at the end we will set $\breve{R}^a_{\phantom{a}bcd} = 0$ to recover the formula applicable here. According to (\ref{gamgra-trans}), we have
\begin{eqnarray}\label{inf gamgra}
\gamgra_a\gamma_{bc} &=& 0, \nonumber \\
{\gamgra'}_a\left(\gamma_{bc} +\mathcal{L}_\xi \gamma_{bc}\right)&=&0.
\end{eqnarray}
Any two torsionless derivative operators can be related by a symmetric connection; thus, in the same way one might write the figurative relation 
``$\ \nabla = \partial + \Gamma \ $'' to define the GR Christoffel symbols, we write ``$\ \gamgra' = \gamgra + E \ $'' to define a connection $E^a_{\phantom{a}bc} = E^a_{\phantom{a}cb}$ between $\gamgra'_a$ and  $\gamgra_a$. By continuity $E^a_{\phantom{a}bc}$ must be at least first order in $\xi$, so (\ref{inf gamgra}) becomes:
\begin{eqnarray}\label{inf gamgra2}
&&\gamgra_a\left(\gamma_{bc} + 2 \gamma_{d(b}\gamgra_{c)}\xi^d\right) - 2\gamma_{d(b}E^d_{\phantom{d}c)a}=0, \nonumber \\
\Rightarrow\quad &&E_{(bc)a} = \left(\gamgra_a\gamgra_{(c}\xi_{b)}\right).
\end{eqnarray}
However, because $E_{abc}$ is symmetric in its last two indices,
\begin{eqnarray}
 && E_{(ab)c} +E_{(ac)b} - E_{(bc)a} = E_{abc}, \nonumber\\
\Rightarrow\quad &&E^{a}_{\phantom{a}bc} = \gamma^{ae}\left(E_{(eb)c} +E_{(ec)b} - E_{(bc)e}\right).\nonumber
\end{eqnarray}
Substituting (\ref{inf gamgra2}) into the right-hand-side and reorganising the derivatives, we find
\begin{eqnarray}
E^{a}_{\phantom{a}bc} &=& \gamgra_{(b}\gamgra_{c)}\xi^a + \gamma^{ae}\left(\gamgra_{[b}\gamgra_{e]}\xi_c + \gamgra_{[c}\gamgra_{e]}\xi_b\right).\nonumber
\end{eqnarray}
Finally, using the defining property of the Riemann tensor, $\gamgra_{[a}\gamgra_{b]}\xi_c = - \half \breve{R}^d_{\phantom{d}cab}\xi_d$, we arrive at the following compact formula:
\begin{eqnarray}
E^{a}_{\phantom{a}bc} &=& \left(\delta^a_d\gamgra_{(b}\gamgra_{c)} - \breve{R}^a_{\phantom{a}(bc)d}\right)\xi^d.
\end{eqnarray}
In the case where $\gamma^{ab}$ is flat, this becomes
\begin{eqnarray}
E^{a}_{\phantom{a}bc} &=& \xi^a_{\phantom{a} \gamsem b \gamsem c}.
\end{eqnarray}

\bibliography{BG_tensor}
\enddocument